\documentclass[journal]{IEEEtran}
\usepackage{graphicx, subfigure}
\usepackage{amsmath}
\usepackage[colorlinks,linkcolor=blue]{hyperref}
\usepackage{cite}

\ifCLASSINFOpdf
\else
\fi
\begin{document}
\title{sDAC---Semantic Digital Analog Converter for Semantic Communications}

\author{Zhicheng~Bao,~Chen~Dong,~Xiaodong~Xu,~\IEEEmembership{Senior~Member,~IEEE}% <-this % stops a space
% \thanks{M. Shell was with the Department
% of Electrical and Computer Engineering, Georgia Institute of Technology, Atlanta,
% GA, 30332 USA e-mail: (see http://www.michaelshell.org/contact.html).}% <-this % stops a space
% \thanks{J. Doe and J. Doe are with Anonymous University.}% <-this % stops a space
\thanks{This work is supported by the National Key R\&D Program of China under Grant 2022YFB2902100.}

\thanks{Zhicheng Bao Chen Dong, Xiaodong Xu, and Ping Zhang are with the State Key Laboratory
of Networking and Switching Technology, Beijing University of Posts and Telecommunications, Beijing 100876, China. The corresponding author is Chen Dong (e-mail: zhicheng\_bao@bupt.edu.cn; dongchen@bupt.edu.cn; xuxiaodong@bupt.edu.cn; pzhang@bupt.edu.cn).}

% \thanks{Cong Li is with the Academy of Space Electronic Information Technology, Xi’an, 710100, China (e-mail: lic1@cast504.com).}
}


\markboth{Journal of \LaTeX\ Class Files,~Vol.~XX, No.~XX, XXX~2023}%
{Shell \MakeLowercase{\textit{et al.}}: Bare Demo of IEEEtran.cls for IEEE Journals}



\maketitle

\begin{abstract}
In this paper, we propose a novel semantic digital analog converter (sDAC) for the compatibility of semantic communications and digital communications. Most of the current semantic communication systems are based on the analog modulations, ignoring their incorporation with digital communication systems, which are more common in practice. In fact, quantization methods in traditional communication systems are not appropriate for use in the era of semantic communication as these methods do not consider the semantic information inside symbols. In this case, any bit flip caused by channel noise can lead to a great performance drop. To address this challenge, sDAC is proposed. It is a simple yet efficient and generative module used to realize digital and analog bi-directional conversion. On the transmitter side, continuous values from the encoder are converted to binary bits and then can be modulated by any existing methods. After transmitting through the noisy channel, these bits get demodulated by paired methods and converted back to continuous values for further semantic decoding. The whole progress does not depend on any specific semantic model, modulation methods, or channel conditions. In the experiment section, the performance of sDAC is tested across different semantic models, semantic tasks, modulation methods, channel conditions and quantization orders. Test results show that the proposed sDAC has great generative properties and channel robustness. 
% Relative code is released at \href{https://github.com/Azul-9/MDMA_video_communication}{https://github.com/Azul-9/MDMA\_video\_communication}.
\end{abstract}


% \fi
\IEEEpeerreviewmaketitle

\section{Introduction}

\begin{figure*}
  \centerline{\includegraphics[width=1\textwidth]{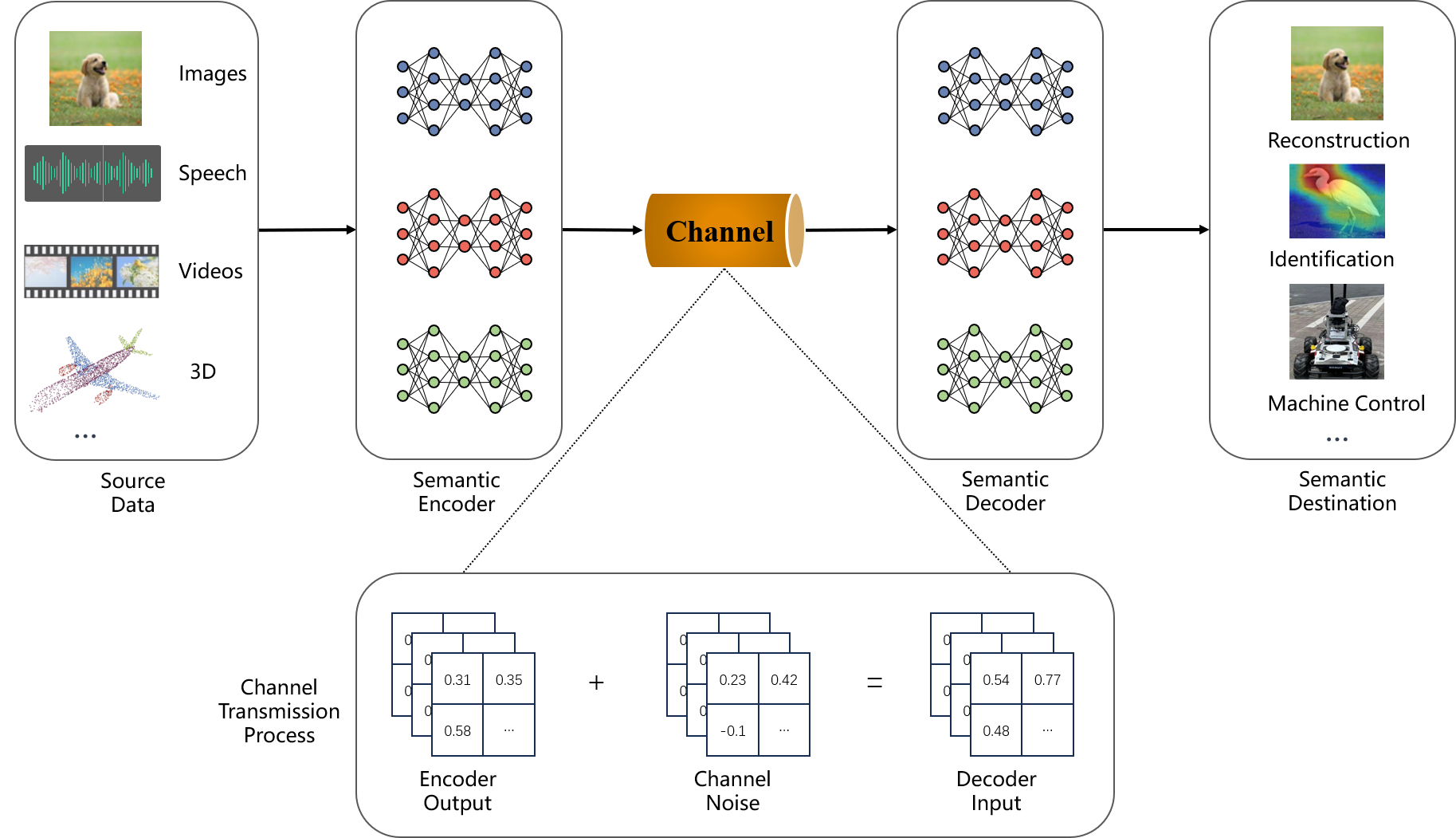}}
  \caption{An abstract common semantic communication framework. Source data of various modalities get encoded by various semantic encoders and output continuous-valued data. These data are added with analog channel noise to simulate wireless noisy channels. At the receiver, these data get decoded by various semantic decoders for various semantic destinations, including traditional reconstruction and rising semantic tasks.
  \label{semantic_framework}}
  \end{figure*}
  
\IEEEPARstart{S}{emantic} communications are evolving as a beyond-Shannon-type communication paradigm. It is expected to be one of the potential technologies for future 6G wireless systems. \cite{9679803, Niu2022APS} Different from the traditional technical-level communication system, the semantic communication system focuses on the conveying of semantic information at the semantic level or the performance of a specific task at the effectiveness level. To be short, semantic communications can improve communication efficiency and performance especially under limited communication resources for its ability of high coding gain and channel robustness.

Most of the existing communication systems use source coding and channel coding separately. Source coding applies various algorithms according to the data modal to remove redundancies and then convert them into bit sequences. After that, channel coding, like low-density parity-check (LDPC) coding, \cite{Richardson2018DesignOL} adds parity check bits to the code to improve communication robustness. This scheme has lots of advantages and has been applied widely in the past few decades. However, its limitation starts to emerge when the demand for higher communication quality under lower communication resources costs increases. The cliff effect, for example, affects communication performance greatly under worse channel conditions than usual. Meanwhile, as the separated scheme only optimizes their performance individually, the joint scheme is expected to have a higher coding gain when taking more communication processes into consideration.

Therefore, joint source-channel coding (JSCC), which combines source coding and channel coding, was studied in \cite{Fresia2010JointSA, Guyader2001JointST}. Based on this idea, with the development of deep neural networks (DNNs), semantic communications build up their semantic codecs for JSCC. Works in \cite{Xie2020DeepLE, Bourtsoulatze2019DeepJS, Kurka2019DeepJSCCfDJ, Yang2021DeepJS, 10477917, Dong2023SemanticCS, 10464666, liu2023semantic, 8461983} study the semantic JSCC communication system of different source data. They build a pair of semantic codecs to realize the reconstruction of texts, speech, images, videos and 3D point cloud. By using DNNs, source data can be directly encoded as continuous-valued symbols to be transmitted through a noisy wireless channel instead of conducting source coding and channel coding independently. Such a system not only shows great robustness to the cliff effect but also improves the efficiency of the communication system, and its performance exceeds the traditional structure. Apart from the traditional end-to-end data transmission semantic communication system, task-oriented semantic communication systems are rising. It transmits only the relevant semantic information of a message instead of the whole message to achieve a specific task. It reduces latency, compresses the data, and is more robust when it comes to channel noise. Following that, the authors in \cite{weng2024robust} propose a task-oriented semantic transmission system to achieve speech-to-text translation. The original source of speech is encoded and transmitted through a wireless channel. At the receiver, semantic data gets decoded into text data instead of recovering speech. It ensures that only the task-related semantic features are extracted, improving communication efficiency. Similarly, works in \cite{10445833} propose a general semantic communication framework for a wide range of control tasks, which balances communication efficiency and task performance well. Works in \cite{Wang2021DeepJS, Jankowski2020WirelessIR, Lo2023CollaborativeSC} study the multi-task architecture to further improve semantic communication efficiency.

To sum up, in many of the current semantic communication systems, as mentioned above, source data of different modalities get encoded by the semantic encoder and the continuous-valued output gets transmitted directly through the noisy channel. During transmission, analog noise like additive white Gaussian noise (AWGN) is added to the encoder's output. The semantic decoder receives this distorted and continuous-valued signal and conducts the following task-oriented steps. The whole transmission process can be abstracted to Fig.~\ref{semantic_framework}. Source data like images, speech, videos, 3D cloud points, and more are processed by the corresponding semantic encoder and transformed into a transmission-ready signal. The corresponding semantic decoder deals with these analog signals and outputs different formats of results according to their tasks. Data at both sides of the channel is not quantized and then modulated in the form of analog. However, modern communication systems are mostly based on digital bits, and many high-efficiency digital modulation technologies have been developed. It makes it hard for semantic communication systems to be compatible with the modern digital communication system. Furthermore, it raises a doubt: Is semantic communication still cliff effect robustness after quantized into discrete bits?

\section{Related works}
\subsection{Semantic digital communication}
Many researchers have proposed their ideas to address the problems mentioned above. The authors in \cite{9838671} propose an end-to-end communication framework to investigate the effects of constraining the channel input to a predefined constellation. By mapping the output of the encoder to the constellation, this system integrates the digital modulation process into the end-to-end communication. Works at \cite{10304507} designs an image semantic communication system to restrain the output of the encoder based on the sub and super constellations. Similarly, works at \cite{10495330, 10159007} integrate digital modulation into the system and map data to the corresponding constellation. Although these works solve the problem of compatibility with the digital modulation, they are bound with a specific modulation method, and any changes in the modulation will result in the re-training of the entire end-to-end system.

Different from the ideas mentioned above, works at \cite{10200355} propose a non-linear quantization method. With trainable quantization levels and sigmoid function, this system can map the dynamic range of float points to the discrete numbers and then modulate 0-1 bits digitally. Works at \cite{Xie2022RobustIB} propose an end-to-end task-oriented semantic communication system with digital modulation to realize inference at the receiver. Works at \cite{liu2024ofdmbased} proposed a system which integrated with OFDM and then quantizing data into OFDM symbols. The authors at \cite{park2024joint} propose a quantization method for channel-adaptive digital semantic communications that provide robustness and flexibility against diverse channel conditions and modulation schemes. Similarly, it uses the sigmoid function to transform the output of the encoder into the range from 0 to 1. These transformed data are interpreted as the probability of the modulated bit being 1. Furthermore, they harness a stochastic model to abstract the combined effects of digital modulation, noisy channel, equalization and demodulation to improve semantic codec generality. After that, a channel-adaptive modulation strategy is proposed to modulate these discrete data according to the channel condition. These works release the binding of digital modulation and semantic codec and have great compatibility with digital communication. However, these works lack generality and only apply to a specific communication framework. It limits the application of the quantization algorithm. Meanwhile, bit decisions that ignore semantic information inside data may affect end-to-end communication performance.

\subsection{Motivations and contributions}
Summarizing the works mentioned above, we can classify them into two categories: constellation-based and quantization-based. The former maps the output of the encoder into constellations directly to avoid infinite data range. However, this method makes the entire communication system and modulation scheme highly coupled. Even a simple change of modulation order from 16QAM to 64QAM, for example, will lead to the re-training of the semantic codec. The latter quantizes the continuous-valued encoder's output into discrete numbers. After that, the following process is the same as the traditional digital modulation communication. It decouples the digital modulation system and semantic communication system. However, the quantization algorithm is manually designed and traditional quantization methods do not consider the semantic information in data. Even the same bit error rate (BER) can result in distinct communication performance. For instance, a value $0$ output by the encoder can be transformed into $000$ through binary. After demodulating, it may become $100$ or $001$, which represents $4$ and $1$ separately. In the view of analog semantic communication, the number $4$ contains much more channel noise energy than the number $1$ even if they have the same BER, and the former will affect communication performance more evidently. More details will be discussed in the following section.

Therefore, based on the discussion above, we hope to propose a novel quantization algorithm and module which has the following properties:
\begin{itemize}
    \item It is a generalized module and can be embedded into a large amount of analog semantic communication systems without the need to modify the original system architecture.
    \item It does not depend on any specific digital modulation methods and is compatible with nearly any existing digital modulation without the need to re-train semantic codec.
    \item It takes semantic information into consideration and is able to accommodate different source data, channel conditions and semantic tasks.
    \item It has a flexible quantization order adjustment to meet demands between the trade-off of communication performance and costs.
\end{itemize}

To be specific, the contribution of this paper can be summarized as the following:

(1) sDAC module: A novel module named semantic digital analog converter (sDAC) is proposed to realize the bi-directional conversion of digital discrete numbers and analog continuous values. It shows great robustness on channel noise and digital modulation. It also has a great generality on a large range of semantic communication systems and tasks. To the best of the author's knowledge, this is the first generalized digital-analog bidirectional converter that is proposed for semantic communication systems.

(2) sDAC discrete training framework: A novel end-to-end discrete training framework and corresponding algorithms are proposed for sDAC. It enables sDAC to converge efficiently and robustly to a wide range of channel conditions. In addition, a flexible quantization order structure is proposed for the trade-off between communication quality and costs.

(3) Average semantic error (ASE) metric: For the measurement of the performance data of sDAC, a new metric named average semantic error (ASE) is proposed. It calculates the mean semantic distance between the original semantic symbols and the distorted ones, considering that the distortion is caused by quantization, channel noise or other processes. It is a more generalized metric that is allied with actual semantic communication performance.

(4) Performance validation: The performance of the proposed sDAC module is verified across different semantic communication systems, semantic tasks, modulation methods and channel conditions. As shown in the test results, the sDAC is generalized in many semantic communication systems and robust to channel distortions.

The rest of this paper is arranged as follows. In the next section III, the overall system model, including the end-to-end communication process and where the sDAC is located, are introduced. Then, sDAC architecture and corresponding algorithms and metrics are detailed in section IV. The experiment part will be shown in section V, and section VI concludes this paper. 

 

% \subsection{Subsection Heading Here}
% Subsection text here.

%\IEEEpubidadjcol

% \subsubsection{Subsubsection Heading Here}
% Subsubsection text here.

\section{System model}
In this section, the system model of the sDAC-based semantic communication system will be presented first. Then, the discrete communication channel will be specified. Following that, the whole communication process will be introduced. In the end, the generalized optimization goal of it will be derived.

\begin{figure*}
  \centerline{\includegraphics[width=1\textwidth]{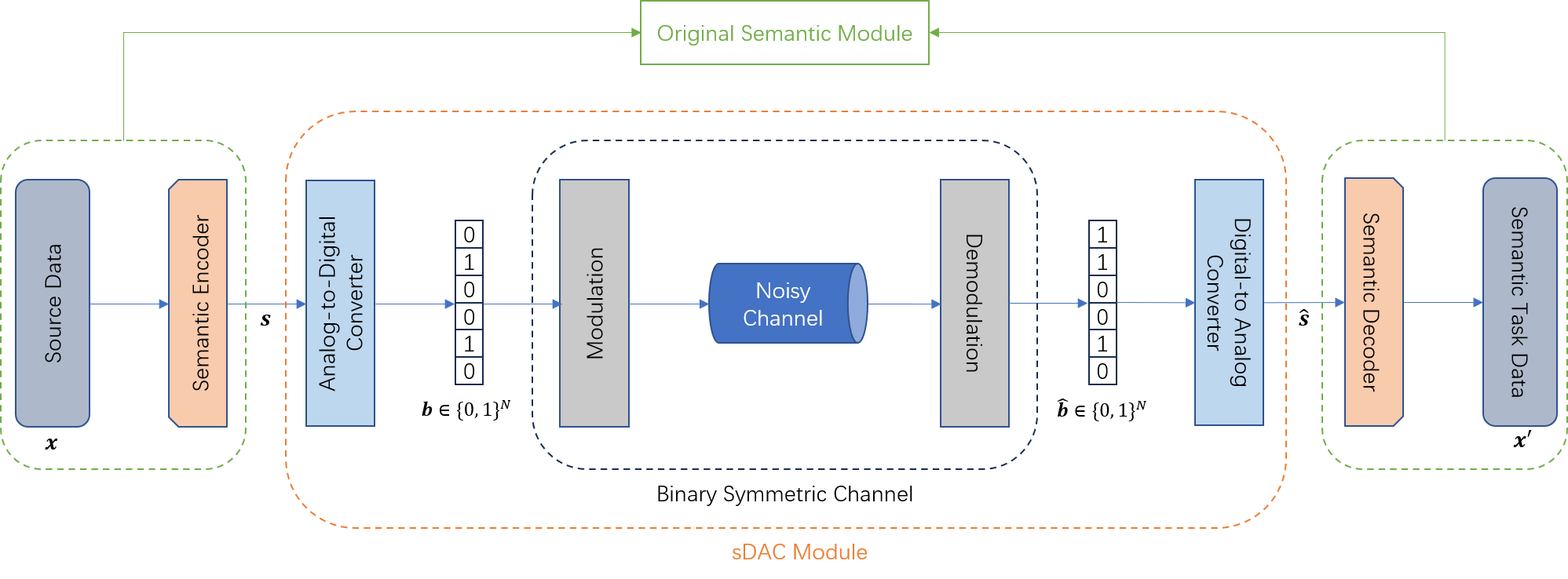}}
  \caption{The illustration of the sDAC-based semantic communication system. It is composed of original semantic modules, DAC modules and binary symmetric channels. The output of the semantic encoder $s$ gets quantized by an analog-to-digital converter and then transmitted through a BSC to simulate the effect caused by intermediate processes, including modulation, equalization, etc. At the receiver, the process above gets inverted by a digital-to-analog converter to output $\hat{s}$. The process above does not modify any module in the original semantic communication system or digital traditional communication system, ensuring it is a great generalization.
  \label{system model}}
  \end{figure*}

\subsection{sDAC-based semantic communication system}
The illustration of the sDAC-based semantic communication system is presented in Fig.~\ref{system model}. It can be divided into three parts: original semantic module, sDAC module and binary symmetric channel. The original semantic module is an abstract semantic communication framework of the generalized communication process. It can work properly like before if removing the sDAC module, which has been shown in Fig.~\ref{semantic_framework}. The sDAC module is embedded into the original semantic module for digital and analog bi-directional transform. It consists of a pair of analog-to-digital converter and digital-to-analog converter. Just as its name implies, the analog-to-digital converter transforms the output of the semantic encoder into discrete bits $b$, and the digital-to-analog converter transforms discrete bits $b$ back into continuous value for the semantic decoder. Attention that the whole architecture \textbf{do not} modify any of the original semantic modules, enabling it a great compatibility on many of the semantic communication systems. The binary symmetric channel (BSC) is a channel model that jointly abstracts traditional modulation methods and other related operations. In the end, it presents a stochastic bit flap phenomenon to the sDAC, enabling it to be robust to different modulation orders and channel conditions. Namely, from the view of the original semantic module, its forward communication path does not get modified. From the sDAC module's perspective, it does not care about the change in the traditional communication process. This decoupled and abstract architecture gives the sDAC great generality.

\subsection{Discrete communication channel}
To facilitate the end-to-end discrete communication process, it is crucial to take all of the processes in traditional digital communication systems into consideration. However, realizing that is not only complicated but also makes the whole communication process not differentiable for the nonlinearity of digital modulation. In light of this, we introduce the BSC channel, which has been adopted widely in traditional communication theory. In which case, a binary bits $b_i\in\{0, 1\}$ is stochastically transformed into another random bits $\hat{b_i}\in\{0, 1\}$ according to a transition probability $p$. This stochastic transformation is used to model the relationship between the analog-to-digital converter's output and the digital-to-analog converter's input. This conditional distribution can be represented as 
\begin{equation}
    p_{BSC}(\hat{b_i}|b_i)=\begin{cases}
        1-P,&\text{if } \hat{b_i}=b_i , \\
        P,&\text{else } ,
    \end{cases}
\end{equation}
where $P$ is the transition probability, namely the bit-flip probability in the context of BSC. 

One of the key advantages of introducing the BSC channel is that it abstracts the intermediate communication process, including digital modulation, fading channel, channel equalization, digital demodulation and so on. It enables a differential computation during communication forward path and a high generality for sDAC. It transforms the complicated digital communication process into a simple bit-flip process, ignoring technology details inside the digital communication and making it possible for sDAC to be compatible with any digital communication system.

\subsection{End-to-end communication}
This subsection gives a one-by-one description of the end-to-end communication process. Assuming that there is a source data $x$ to be transmitted through a wireless channel. It can be any type of data as long as there is a corresponding semantic encoder $f_\theta(\cdot)$ to process. The output of the semantic encoder is $s$ and then sent into the analog-to-digital converter $f_{ad}(\cdot)$. After that, binary bit sequence $b$ get transmitted through the BSC $f_{BSC}(\cdot)$, where $b\in\{0, 1\}^N$ and $N$ is the length of the bit sequence. It should be paid attention that $N$ is related to the quantization order $q$, which is detailed in section IV. At the receiver end, distorted bit sequence $\hat{b}\in\{0, 1\}^N$ gets transformed back to analog continuous value $\hat{s}$ by digital-to-analog converter $f_{da}(\cdot)$. In the end, the recovered semantic signal gets processed by semantic decoder $f_\psi(\cdot)$ and outputs semantic task data $x'$. Attention that $x'$ is not necessarily the reconstruction of $x$ in some task-oriented communication scenes, which is presented in section V. To sum up, the whole communication link of the sDAC-based semantic communication system can be formulated as 
\begin{equation}
    x\xrightarrow{f_\theta(\cdot)}s\xrightarrow{f_{ad}(\cdot)}b\xrightarrow{f_{BSC}(\cdot)}\hat{b}\xrightarrow{f_{da}(\cdot)}\hat{s}\xrightarrow{f_\psi(\cdot)}x'.
\end{equation}

\subsection{Optimization goal}
The optimization goal of the sDAC is to maximize the performance of the original semantic communication system in the end-to-end discrete digital link. Without the loss of generality, the target function to be optimized by the original semantic communication system is denoted as $L_{ori}(\theta, \psi)$. It is used to optimize corresponding semantic tasks such as image reconstruction, image classification and so on. Corresponding to it, the target function of sDAC is denoted as $L_{sDAC}(ad, da)$. It is used to minimize the performance drop caused by discrete digital communication. Therefore, the overall loss function of the sDAC-based semantic communication system can be formulated as 
\begin{equation}
    L=L_{ori}(\theta, \psi)+\lambda \cdot L_{sDAC}(ad, da),
    \label{loss eq}
\end{equation}
where $\lambda$ is a hyper-parameter to control the trade-off between the quantization loss and original semantic task loss. Details about the $L_{sDAC}(ad, da)$ are presented in section IV.

\section{sDAC module and proposed methods}
This section presents a detailed explanation of the structure, implementation, and algorithms of DAC. Meanwhile, the corresponding training framework designed for a sDAC-based semantic communication system is detailed. In addition, a metric named average semantic error (ASE) is proposed to explain and quantify the mechanism of sDAC.

\subsection{The framework of sDAC}
\begin{figure*}
  \centerline{\includegraphics[width=1\textwidth]{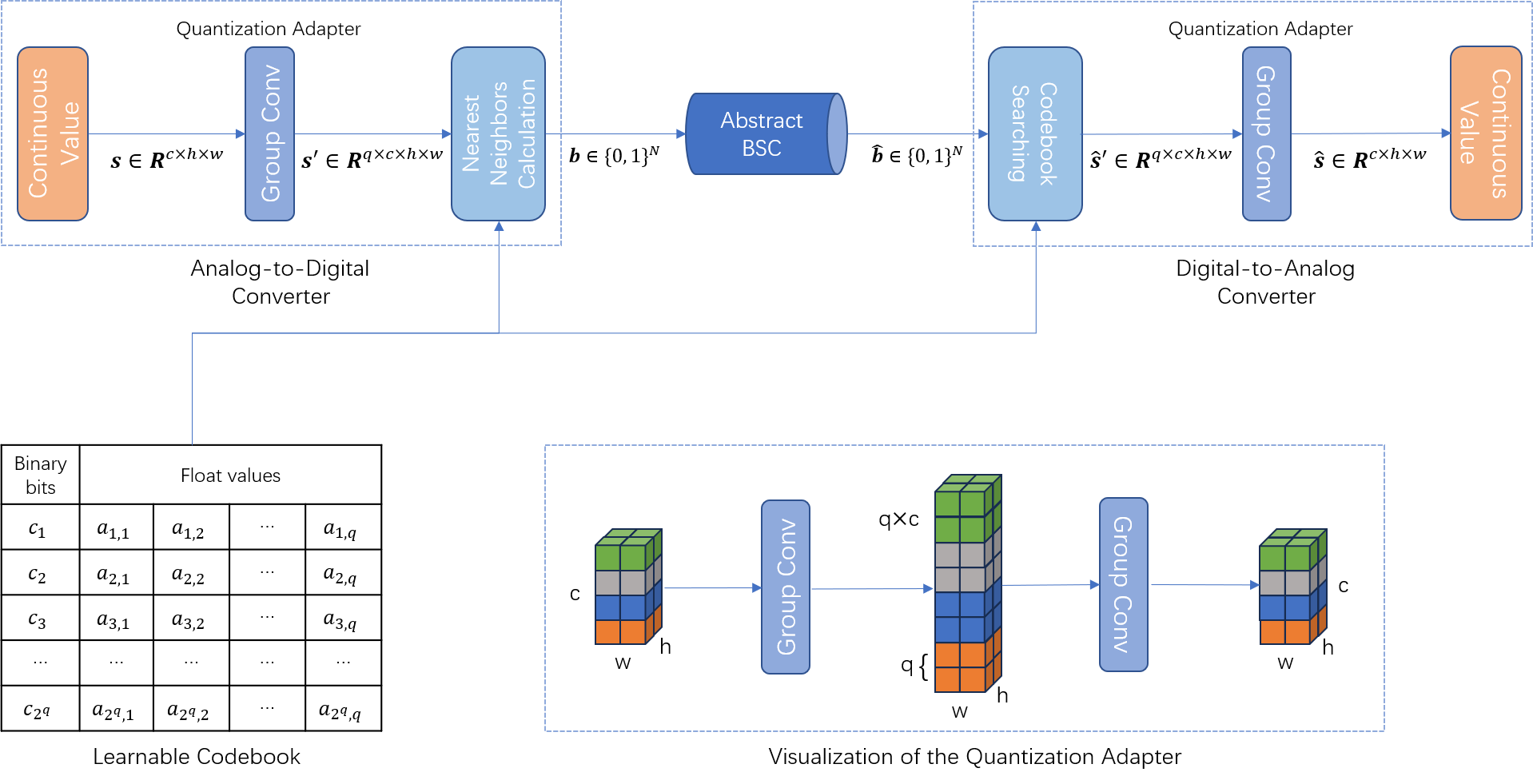}}
  \caption{The internal structure of sDAC. Analog-to-digital converter transforms continuous values into binary bits through the nearest neighbours calculation between the learnable codebook and the output of the quantization adapter, which is a group convolution to map one single data to $q$ symbols. The digital-to-analog converter realizes the inverse process of combining binary bits back to continuous values for further semantic tasks. The end-to-end communication process is powered by the abstract BSC to simulate the digital communication process.
  \label{sDAC model}}
\end{figure*}
To further explain the mechanism of sDAC, its structure is visualized in Fig.~\ref{sDAC model}. It can be divided into a pair of converters, a shared learnable codebook and an abstract BSC channel.

At the analog-to-digital converter side, the output of the semantic encoder, which is a vector of continuous value, is denoted as $s\in R^{c\times h\times w}$. Here, we define quantization order $q$ to imply how many bits should be used to represent a continuous value. After the processing of the quantization adapter, which is a group convolution calculation $f_g(\cdot)$, the original signal $s$ is transformed into $s'\in R^{q\times c\times h\times w}$. Attention that $s'$ is still a continuous-value vector. Group convolution \cite{Krizhevsky2012ImageNetCW} is a kind of special convolution calculation where data at $c$ dimension are divided into groups, and during calculation, only data from the same group get calculated, while normal convolution considers all of the data at $c$ dimension. In our sDAC, the quantization adapter is used to expand original single data to $q$ continuous-value data for further process. In theory, the $q$ expanded data is only related to one original data, and normal convolution calculation, which considers all of the data at $c$ dimension, will introduce interference during expansion. Therefore, group convolution is a better choice for cancelling interference. The visualization of it is shown in the right-down corner of Fig.~\ref{sDAC model}. Data that is divided into the same group is painted in the same colour. 

After that, $s'$ is sent to the nearest neighbors calculation module to output binary bit sequences. To achieve that, a learnable codebook $E$ is defined to maintain a bi-directional map between binary bit sequences $C$ and corresponding float values $A$. It is denoted as 
\begin{equation}
    E=[a_1, a_2, ..., a_{2^q}].
\end{equation}
It means that a continuous value can be quantized into $2^q$ different bit sequences if using $q$ binary bits. For a specific entry $c_i$ and $a_i$, this map is denoted as 
\begin{equation}
    \begin{split}
        \text{MAP}(c_i, a_i)=A\xrightarrow{\text{Euclid distance}} C,\\
       \text{with}\{c_i, a_i | c_i \in \{0, 1 \}^q, a_i \in R^q\}
    \end{split}
    \label{eq1}
\end{equation}
It is a bijection where a sequence of continuous-value vector $a_{i}\in R^q$ can only be mapped to a unique binary sequence $c_i$, and vice versa. Here, Euclid distance $D$ is introduced to minimize the loss generated from quantization. It is defined as 
\begin{equation}
    D=\sum_{j=1}^{q}(s'_{i, j} - a_{i, j}), \text{with }a_{i, j},s'_{i, j}\in R.
    \label{euclid}
\end{equation}
For every grouped data $s'_i \in s'$, its vector component $s'_{i, j}$ gets calculated with all of the continuous values in the codebook and output Euclid distance matrix $m_e \in R^{2^q}$. It indicates that $s'$ should be quantized to which float value sequences to minimize quantization error. Therefore, Eq.~\ref{eq1} can be further detailed as 
\begin{equation}
    a_i \xrightarrow{} c_i, i = \text{arg min}D.
    \label{argmin}
\end{equation}
So far, the original continuous-value signal $s$ has been transformed to binary bit sequences $b \in \{0, 1 \}^N$ and then transmitted through the abstract BSC. Here $N$ and $q$ satisfying the following equation:
\begin{equation}
    N=(c\times h\times w)\times q.
\end{equation}
It means that for every single data $s_i$, it gets expanded to $q$ continuous values and further quantized to $q$ binary bits $c_i$, where $s_i \in R$.

On the digital-to-analog converter side, its process is approximately the reverse of the analog-to-digital side. After receiving the distorted binary bit sequences $\hat{b} \in \{0, 1 \}^N$, the codebook searching module will map $\hat{s}$ back to continuous-value data according to a learnable codebook with every $q$ binary bits. It is a table-searching operation without any calculation, slightly releasing the computation burden at the receiver. After that, $\hat{s}' \in R^{q\times c\times h\times w}$ gets transformed back to $\hat{s} \in R ^{c\times h\times w}$ by group convolution $f_{\hat{g}}(\cdot)$. At this point, sDAC has finished its process, and the following operation is conducted by the semantic decoder for various semantic tasks.

\subsection{sDAC discrete training framework}
Regarding the forward calculation path of the sDAC-based semantic communication system, one of the tough problems caused by the digital communication system is the differential property introduced by binary bits. It makes it hard to optimize end-to-end communication systems as the gradient cannot be calculated through the chain derivative rule. To solve this issue, a discrete training framework designed for the sDAC-based semantic communication system is proposed, assuming that the original semantic communication system is derivable.

The above forward calculation path can be formulated as 
\begin{equation}
    s\xrightarrow{f_g(\cdot)}s'\xrightarrow{\text{nearest neighbors}}b\xrightarrow{\text{BSC}}\hat{b}\xrightarrow{\text{replacing}}\hat{s}'\xrightarrow{f_{\hat{g}}(\cdot)}\hat{s}.
\end{equation}
The non-derivable problem occurs at nearest neighbours, BSC and replacing binary with continuous-value data process. To formula that, we first directly denote the latter form in the loss function mentioned in Eq.~\ref{loss eq} as
\begin{equation}
    L_{sDAC}(ad, da)=||f_{\hat{g}}(s') - s||_2^2.
    \label{eq2}
\end{equation}
In the end-to-end communication system, the formula above is derivable as it contains no quantization or binarization. However, the actual input of $f_{\hat{g}}(\cdot)$ is $\hat{s}'$ instead of $s'$, it is obviously not the optimization goal of sDAC. In theory, Eq.~\ref{eq2} should be overwritten as 
\begin{equation}
    L_{sDAC}(ad, da)=||f_{\hat{g}}(\hat{s}') - s||_2^2.
    \label{eq3}
\end{equation}
Nevertheless, it introduces a new problem that directly selecting $a_i$ according to Eq.~\ref{argmin} will stop gradient propagation, preventing the update of $f_{\hat{g}}$ and $f_g$. Therefore, follow the work in \cite{Oord2017NeuralDR}, Eq.~\ref{eq3} is rewritten as 
\begin{equation}
    L_{sDAC}(ad, da)=||f_{\hat{g}}(s' + \text{sg}[\hat{s}' - s']) - s||_2^2,
    \label{eq4}
\end{equation}
where sg means stop gradient, treating these variables as independent values that are not related to other variables, namely erasing their gradients. From the view of the chain derivative rule, Eq.~\ref{eq4} is equivalent to Eq.~\ref{eq2} as they generate the same gradient for the model update. Meanwhile, from the view of the output value of the loss function, Eq.~\ref{eq4} is equivalent to Eq.~\ref{eq3} as they generate the same loss for the chain derivative rule. That is to say, we jointly optimize the loss function of sDAC and update $f_{\hat{g}}$ and $f_g$ by replacing the gradient of $\hat{s}'$ with $s'$. It solves the optimization problem for the quantization adapter. Next, the training algorithms of the learnable codebook will be presented to realize the end-to-end training of the sDAC-based semantic communication system interactively.

Considering that the output of the semantic encoder is quantized to $a_i$ to minimize the quantization error, it is a better choice to update the codebook $E$ along with the quantization adapter during training. In theory, the closer $s'$ and $a_i$ are, the smaller the error indicated by Eq.~\ref{euclid}. Unfortunately, even though the outputs of the loss function Eq.~\ref{eq2} and Eq.~\ref{eq3} are quite small, it still cannot guarantee the decrease of the Euclid distance. Therefore, kl loss, which is used to measure the distance of distribution, is introduced to restrain $s'$ and $\hat{s}'$. It is denoted as 
\begin{equation}
    ||s' - \hat{s}'||_2^2.
    \label{eq5}
\end{equation}
Similar to Eq.~\ref{eq4}, it can be rewritten as 
\begin{equation}
    ||\text{sg}[s'] - \hat{s}'||_2^2 + ||s' -\text{sg}[\hat{s}']||_2^2.
\end{equation}
From the view of the gradient, it is equivalent to Eq.~\ref{eq5}. The former term means to fix $s'$ and update $\hat{s}'$, which is the codebook's output $a_i$, to narrow kl loss. The latter term means fixing the codebook and updating the encoder's output to narrow the kl loss. Considering that updating the codebook is easier than updating $f_g$ to fit their distribution, a pair of hyper-parameters $\alpha$ and $\beta$ are used to control the trade-off between. To this end, the final version of the loss function for sDAC is formulated as 
\begin{equation}
\begin{split}
    L_{sDAC}(ad, da)=\\&||f_{\hat{g}}(s' + \text{sg}[\hat{s}' - s']) - s||_2^2 + \\&\alpha||\text{sg}[s'] - \hat{s}'||_2^2 + \\&\beta||s' -\text{sg}[\hat{s}']||_2^2
    \\&\text{with } \alpha + \beta = 1.
\end{split}
\end{equation}

The last calculation that is not derivable is the BSC model. It stochastically flaps binary bits. In view of that, a simple solution is to transform discrete channel noise caused by a bit-flap operation to continuous channel noise caused by the additive operation. Namely, simulating BSC noise by analog derivable channel noise. It can be modelled as
\begin{equation}
    \hat{b} = b + n, \text{with }n_i\in \{0, 1\}.
\end{equation}
In practice, as $b_i \in \{0, 1\}$, above equation can be directly refined as
\begin{equation}
    \hat{b} = \{\hat{b}_i | 1 - b_i,\}, \text{ with } i \in \{i|p_i < p \}, p_i \sim \mathcal{U}(0, 1),
\end{equation}
where $p$ is the bit-flip probability or bit error rate (BER), $p_i$ is a random number sampled from a standard uniform distribution to indicate if $b_i$ should be flapped. To sum up, this discrete training framework substitutes operations that are not derivable by derivable and equivalent calculations. It makes it possible to optimize the end-to-end communication link for the sDAC-based semantic communication system.

\subsection{ASE metric}
\begin{figure}
  \centerline{\includegraphics[width=0.5\textwidth]{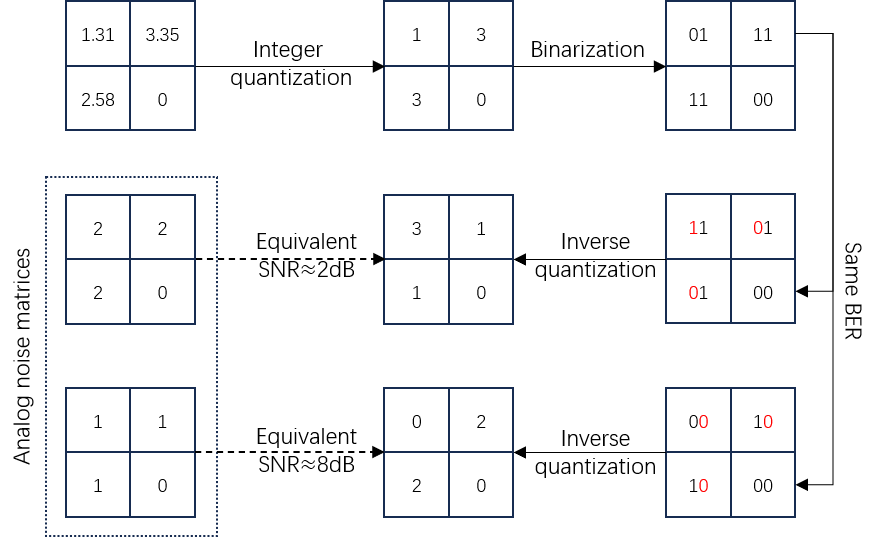}}
  \caption{An illustration of different SNR effects under the same BER with binary quantization. It proves that it is not BER but semantic distance, also known as semantic error, that really counts. It also makes it hard to reflect the end-to-end communication performance with a single metric BER.
  \label{ASE}}
\end{figure}
The traditional communication system focuses on the precision transmission of binary bits, hence proposing a bit error rate (BER) metric to measure the channel condition or communication performance. In the context of semantic communication, this metric cannot fully reflect the end-to-end communication performance, although it is still an appropriate metric for channel noise. One of the reasons for that is BER ignores the semantic information in binary bits and treats all of them equally. However, even the same BER can result in distinctly different performance fluctuations. A simple example is shown in Fig.~\ref{ASE}.

Assuming that there is a continuous-value matrix $m$ to be transmitted. After integer quantization and binarization, $m$ has been transformed into binary-value matrix $m_\text{discrete}$. Through the processing of the digital communication system, the receiver gets distorted matrices $\hat{m}_a$ and $\hat{m}_b$. They have the same BER when compared with $m_\text{discrete}$. In Fig.~\ref{ASE}, BER can be calculated as $\frac{3}{8}$. Then, these matrices get transformed back into integer matrices. Here, the decimal system is used to convert binary bits to integer values. Corresponding analog noise matrices are shown on the left side of the illustration, and accordingly, an equivalent signal-to-noise ratio (SNR) can be calculated to show the effect of the channel noise. It is easy to find out that even though $\hat{m}_a$ and $\hat{m}_b$ have the same BER, their semantic communication performance are quite different, with the SNR changing from 2dB to 8dB. In this case, the higher the quantization order is, the more severe the problem is caused by the bit-flap. In the worst condition, a flapped bit can introduce $2^q$ analog noise to the semantic communication system. Therefore, it is necessary to propose a new metric to measure the noise effect caused by quantization methods and channel conditions.

Considering that the semantic communication system deals with data in analog domain and its performance is nearly inversely proportional to the noise power, we define the average semantic error (ASE) as the following
\begin{equation}
    \text{ASE}=\frac{1}{N}\{\sum_{i=1}^{i=N}[\alpha||a_i - s'_i||_2 + \beta||\hat{s}'_i - a_i||_2 + \gamma||\hat{s}_i - s_i||_2]\}.
    \label{ASE_formula}
\end{equation}
The first term $||a_i - s'_i||_2$ indicates the semantic error caused by kl loss between the output of the quantization adapter and the learnable codebook. Generally speaking, it is related to quantization order $q$, and in theory, the higher $q$ is, the smaller the error will be. However, it has the law of diminishing marginal, as presented in section V. The second term denotes the semantic error caused by channel noise. It calculates the semantic distance in terms of a learnable codebook, which considers semantic information in data. The last term means the end-to-end semantic quantization error in the communication system. It includes all of the errors induced in the end-to-end communication process and represents the overall noise to the original semantic modules. The hyper-parameters $\alpha$, $\beta$, and $\gamma$ are used to control the trade-off between these three terms. It is noted that the ASE metric synthetically measures the error of quantization, channel noise and end-to-end communication. 

\section{Experiments}
In this section, sDAC is validated across different semantic communication systems, semantic tasks, modulation methods, channel conditions and quantization orders. Generally speaking, one semantic communication system is designed for one semantic task. Therefore, we combine the semantic communication system part and the semantic task part together to test sDAC. To be specific, sDAC is tested with LSCI \cite{Dong2023SemanticCS}, MDVSC \cite{10464666}, and PCSC\cite{liu2023semantic} for images, videos and 3D point clouds transmission, respectively. Besides, it is also tested with ResNet \cite{7780459} for performance validation in non-reconstruction semantic communication tasks, such as image classification. In addition, for modulation methods, channel conditions and quantization orders experiments, to relieve the influence of the semantic communication codec, a simple semantic communication link is established independently, which will be shown in the setup subsection. 

To be short, in this section, the experimental setup will be introduced first in subsection $A$. Then, experiment results about different communication conditions and semantic communication tasks will be presented in subsection $B$ and subsection $C$, respectively. An overall conclusion will be summarized in the end.

\subsection{Experimental Setup}
\subsubsection{A basic analog semantic communication link}
\begin{figure*}
  \centerline{\includegraphics[width=1\textwidth]{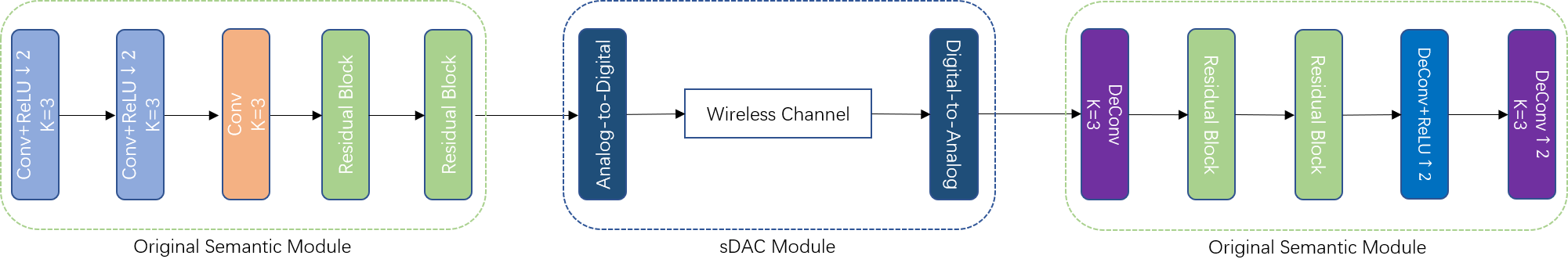}}
  \caption{The illustration of a basic analog semantic communication system. It is implemented to alleviate the influence caused by the semantic codec. Up and down arrows mean 2X up and downsampling. $K$ indicates the kernel size of convolution. This system is based on image source data and for image transmission tasks.
  \label{baseline}}
\end{figure*}
To alleviate the influence caused by the semantic codec and keep the basic communication link the same when testing other scenes at the same time, a basic analog semantic communication link is established for sDAC performance validation. Its structure is shown in Fig.~\ref{baseline}. Up and down arrows in blocks indicate up-sample and down-sample, respectively, $K=3$ means the kernel size of convolution to be 3, and residual block means the module proposed in \cite{7780459}, and sDAC is integrated as the intermediate layer to conduct digital and analog bi-directional converter. This system is designed for image transmission, and the end-to-end communication performance can be measured using the peak signal-to-noise ratio (PSNR) and multi-scale structure similarity (MS-SSIM) \cite{Wang2003MultiscaleSS}. These metrics can be calculated as follows: 
\begin{equation}
    \text{MSE}(X, Y) = \frac{1}{wh}\sum_{i=0}^{w-1}\sum_{j=0}^{h-1}||X(i, j)- Y(i, j)||_2,
\end{equation}
where $m$ and $n$ denote the number of pixels horizontally and vertically.
\begin{equation}
    \text{PSNR}(X, Y) = 10 \cdot \log_{10}(\frac{1}{\text{MSE}}),
\end{equation}
\begin{equation}
\begin{split}
    \text{MS-SSIM}(X, Y) = \\
    [\mathcal{L}_M(X, Y)]^{\alpha_M}\cdot\prod_{j=1}^M[\mathcal{C}_j(X, &Y)\cdot\mathcal{S}_j(X, Y)]^{\alpha j},
\end{split}
\end{equation}
where M means different dimensions, including luminance, contrast, and structure. Therefore, the PSNR metric measures the accuracy of the frame reconstruction, while the MS-SSIM metric evaluates the perceptual quality of the recovered frame. Performance validations conducted in subsection $B$ are all based on this system.

\subsubsection{Implementation details}
The sDAC only has one hyper-parameter $q$ to control the quantization order. It will initialize a $q \times 2^q$ learnable codebook automatically. Regarding the semantic communication system presented in Fig.~\ref{baseline}, number of channels is set to 128 for all of the modules. Batch size and learning rate are set to 64 and $10^{-4}$ separately, and it is trained for 30 epochs with Vimeo-90k \cite{Xue2017VideoEW} datasets. For performance validation, Cityscapes dataset \cite{Cordts2016TheCD} is applied for computing PSNR and MS-SSIM metrics. In addition, unless otherwise specified, all of sDAC is trained under BSC within BER randomly sampled from a uniform distribution: $p \sim \mathcal{U}(0, 0.3)$. The reason for the right boundary of uniform distribution to be 0.3 instead of 0.5 is that if the channel noise is too powerful, sDAC cannot converge to a decent performance as it considers the ability of channel noise robustness in extremely terrible communication conditions, affecting its performance in common conditions. Meanwhile, a BER of 0.3 is approximating -9dB if modulated with BPSK. We believe that setting the maximum of BER to 0.3 is enough to cover most of the communication conditions. 

\subsubsection{Comparison Schemes}
\begin{figure}
  \centerline{\includegraphics[width=0.45\textwidth]{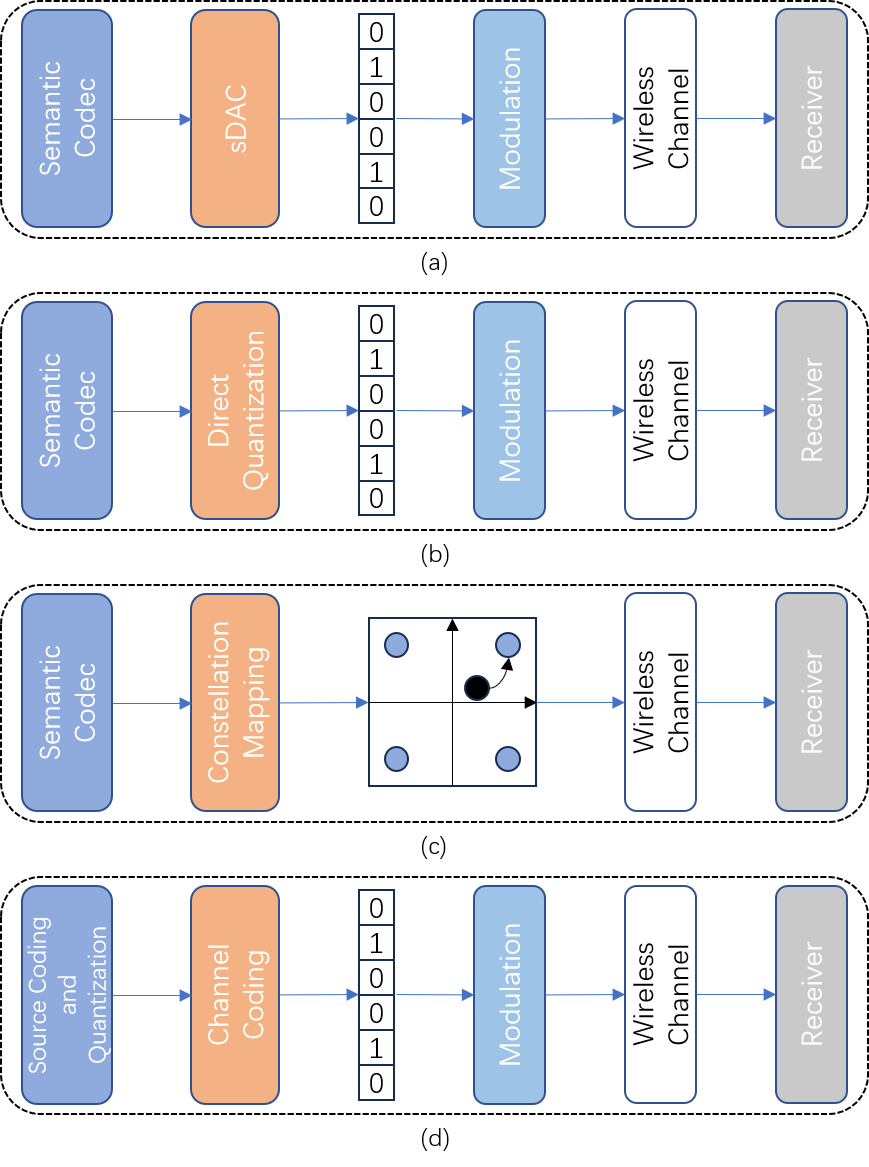}}
  \caption{The illustration of the comparison schemes. Subfigure (a) shows the sDAC-based semantic communication system; it converts continuous-valued data into binary bits without modifying another process in both semantic communication and digital communication. Subfigure (b) shows the direct quantization (DQ) scheme. It also converts continuous values into bits based on the handcrafted decimal conversion without any modification. Subfigure (c) shows the constellation mapping (CM) scheme. It maps continuous values to modulated constellations directly instead of converting them to bits. This scheme has to change both the semantic communication system and the traditional digital system. Subfigure (d) represents the traditional digital communication system based on source and channel split coding.
  \label{comparison}}
\end{figure}
The sDAC-based semantic communication system is compared with the digital semantic communication system and digital traditional communication system. It is noted that the digital semantic communication system is implemented using two schemes: direct quantization (DQ) and constellation mapping (CM). These four schemes are illustrated in Fig.~\ref{comparison} for a clear comparison. As the receiver's process is approximately the inverse of the transmitter, for simplicity, these processes are integrated into the "Receiver" block. Subfigure (a) shows the sDAC-based semantic communication system, which is denoted as \textbf{sDAC} in the following experiments for short. It uses sDAC module to convert continuous-value to binary bits. It nearly does not modify any module in the digital communication system or original semantic communication system. Subfigure (b) shows a similar architecture as subfigure (a) by replacing sDAC with a handcrafted quantization algorithm. It is denoted as \textbf{DQ}. Subfigure (c) shows an end-to-end constellation mapping scheme named \textbf{CM}. This scheme has to modify both the semantic communication system and digital communication system as it is deeply bound by modulation methods. Subfigure (d) shows a traditional digital communication system which does not rely on semantic JSCC codec. It is denoted as \textbf{Traditional}. All of the analog communication systems are trained under AWGN with different SNRs.

In an analog semantic communication system, there is no concept of BER but SNR, as both the output and channel noise are continuous values. Therefore it is necessary to convert SNR to its corresponding BER for its performance tests under digital communication link. However, it is noted that SNR and BER do not have a general conversion formula if digital modulation methods are not specified. In our experiments, M-QAM, QPSK and BPSK modulation methods are applied for performance validation. In this case, for in-phase and orthogonal independent decisions, their symbol error rate (SER) can be formulated as 
\begin{equation}
    P_{\sqrt{M}}=\frac{\sqrt{M} - 1}{\sqrt{M}}\text{erfc}(\sqrt{\frac{d^2}{4N_0}})
\end{equation}
where $d$ is the minimum Euclid distance of constellation, and $N_0$ is the power of noise, bit distribution is assumed to be equiprobable. Based on this, the mean SER of M-QAM can be formulated as 
\begin{equation}
    P_M = 1-(1-P_{\sqrt{M}})^2\approx 2P_{\sqrt{M}}=2\text{erfc}(\sqrt{\frac{d^2}{4N_0}}).
    \label{P_M}
\end{equation}
When using grey code and SER is low, it is assumed that every error symbol has only one error bit, Eq.~\ref{P_M} can be refined to BER as 
\begin{equation}
    P\approx\frac{P_M}{\log_2M}
\end{equation}
Similarly, for BPSK and QPSK, their BER can be calculated as 
\begin{equation}
    P=P_b=P_q=\frac{1}{2}\text{erfc}(\sqrt{\text{SNR}}).
    \label{P}
\end{equation}
Therefore, Eq.~\ref{P_M} and Eq.~\ref{P} are used to establish a connection between BER and SNR.

\subsection{Results of various communication schemes}
\subsubsection{Quantization order and channel conditions}
\begin{figure}
  \centerline{\includegraphics[width=0.5\textwidth]{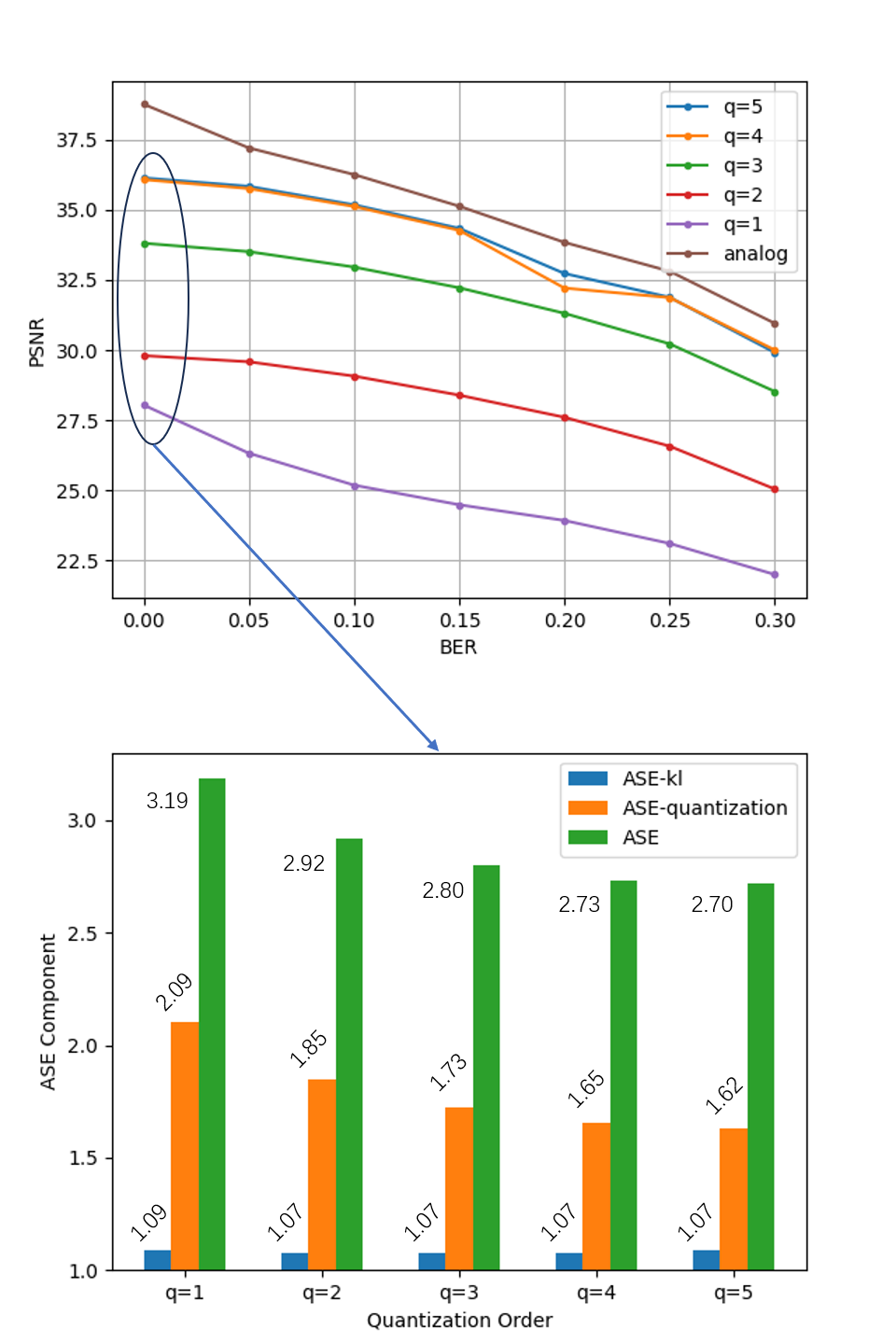}}
  \caption{Results of sDAC at different quantization order under different channel conditions. It can be found that the higher the quantization order is, the better the performance will be. However, it has limitations, and to explain this, quantization error quantification by ASE without the distraction of channel noise is shown below. ASE-kl denotes the quantization error caused by the divergence between the learnable codebook and the output of the semantic encoder. ASE-quantization denotes the end-to-end error caused by the sDAC module.
  \label{quantization}}
\end{figure}
Firstly, the performance of sDAC at different quantization order $q$ under different channel conditions, which is represented by BER, is validated, and results are shown in Fig.~\ref{quantization}. Performance of MS-SSIM is omitted for its similar trend with PSNR. It is easy to find out that with the increasing quantization order $q$, the performance gap with the analog communication system is bridging. However, there exists a limitation. As is shown in Fig.~\ref{quantization}, when $q\leq 3$, the increasing of quantization order can bring evident performance gain. Nevertheless, the performance improvement from 3-bit to 4-bit quantization starts to shrink. Finally, when $q=5$, it nearly cannot bring performance gains in the PSNR metric. This may be caused by the distribution divergence between the learnable codebook and the output of the semantic codec. It indicates that simply increasing the dimension of the learnable codebook cannot infinitely narrow the distribution divergence. To prove this, the ASE metrics of sDAC at different quantization order $q$ are shown at the bottom of Fig.~\ref{quantization}. It shows the value of the ASE metric along with its components of the first term and last term, which are denoted as ASE-kl and ASE-quantization. To remove the effect of channel noise, BER is set to 0 and hyper-parameters in Eq.~\ref{ASE_formula} are set to $\alpha=1, \beta = 0, \gamma=1$. It can be found that the ASE-kl term drops a little bit from 1.09 to 1.07 when increasing the quantization order, indicating that the error caused by the quantization process from the output of the quantization adapter to the learnable codebook is reaching its boundary. Therefore, the performance improvement in the ASE metric is from the ASE-quantization term. Higher quantization order $q$ will reduce quantization noise caused to original semantic modules, thus improving end-to-end communication performance. Based on this conclusion and taking efficiency into consideration, the quantization order $q$ of sDAC in the following experiments is fixed to 4 by default if not specified.

\subsubsection{Modulation methods and channel conditions}
\begin{figure*}
  \centerline{\includegraphics[width=1\textwidth]{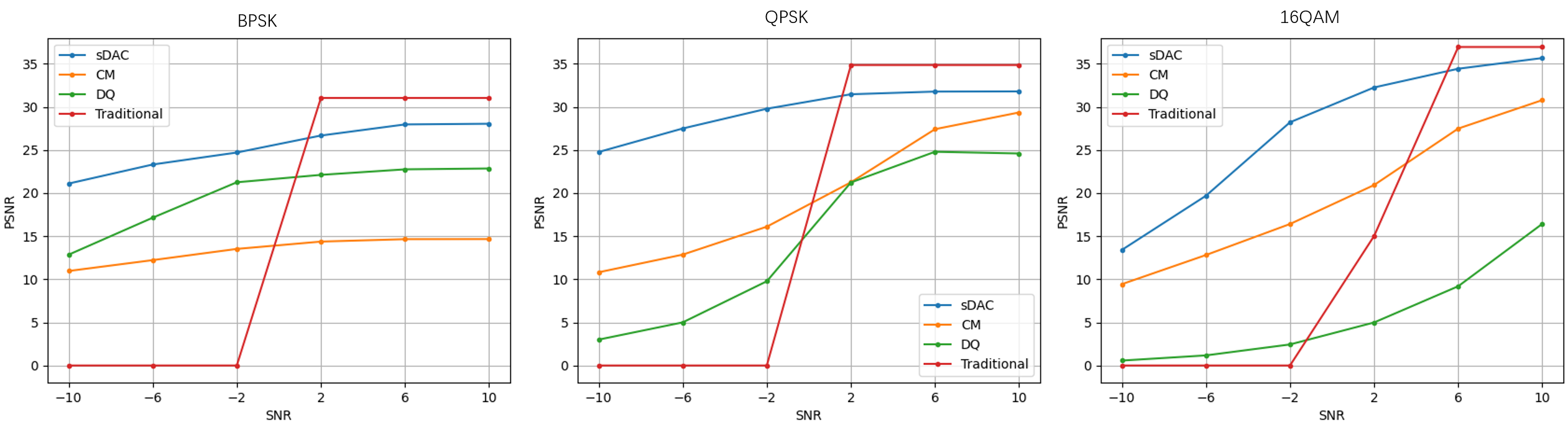}}
  \caption{Results of different modulation methods under different analog channel conditions with different communication schemes. sDAC performs well under low SNR channel conditions and has a small performance gap compared to traditional schemes in high SNR regions. The CM scheme shows a similar trend to that of sDAC. However, it lacks generality and has a performance gap with sDAC. DQ scheme synthetically performs worst for the price of low complexity and is easy to implement. The traditional scheme performs best under decent channel conditions, with an evident cliff effect when facing noisy channels. 
  \label{modulation}}
\end{figure*}
Secondly, the performance of sDAC at different modulation methods under different channel conditions, which is represented by SNR, is validated, and results are shown in Fig.~\ref{modulation}. It is noted that in this subsection, to be compatible with a specific modulation method and communication system, BER is converted to SNR according to Eq.~\ref{P_M} and Eq.~\ref{P}. The traditional scheme is coded with JPEG and 1/4 LDPC. Meanwhile, for the convenience of comparison, the data amount is controlled to be close, and the y-axis is set at the same plotting scale. In all of the digital communication schemes under all of the tested modulation methods, the sDAC-based semantic communication system outperforms other schemes except the traditional scheme under high SNR. With the increasing modulation orders, sDAC and the constellation mapping (CM) scheme share a similar performance improvement trend, but the sDAC has a better result. The traditional scheme achieves better performance than all of the other schemes at high SNR channel conditions. However, it faces a cliff effect earlier at a higher modulation order under the same SNR. The direct quantization (DQ) scheme performs well under low modulation order as these modulation methods have great noise robustness. For the DQ scheme of $q$ quantization order, it can generate $2^q$ semantic error at the worst condition when only one bit gets flipped. Therefore, it outperforms the CM scheme under BPSK as, at this time, the output of the encoder can only be quantized to two constellations, which greatly restricts the semantic encoder. Meanwhile, BER is low enough to ensure the correct transmission of quantized binary bits of the DQ scheme. However, with the increasing modulation order, BER increases rapidly and makes the DQ scheme hardly work, as is shown in the right subfigure of Fig.~\ref{modulation}.

For the comparison under a specific modulation method, it is clear that the traditional scheme performs best under high SNR thanks to the lossless transmission guaranteed by the channel coding LDPC. With the deterioration of channel conditions, LDPC cannot correct all error bits, resulting in a cliff effect. When traditional schemes cannot work properly, the sDAC-based semantic communication system still performs well under low SNR and outperforms all of the other digital schemes. It also replies to the question mentioned in the introduction section: if semantic communication still has cliff effect robustness after getting quantized into discrete bits. It is clear that sDAC-based and CM semantic communication systems show great cliff effect robustness. 

\subsection{Results of various semantic communication tasks}
\begin{figure}
  \centerline{\includegraphics[width=0.5\textwidth]{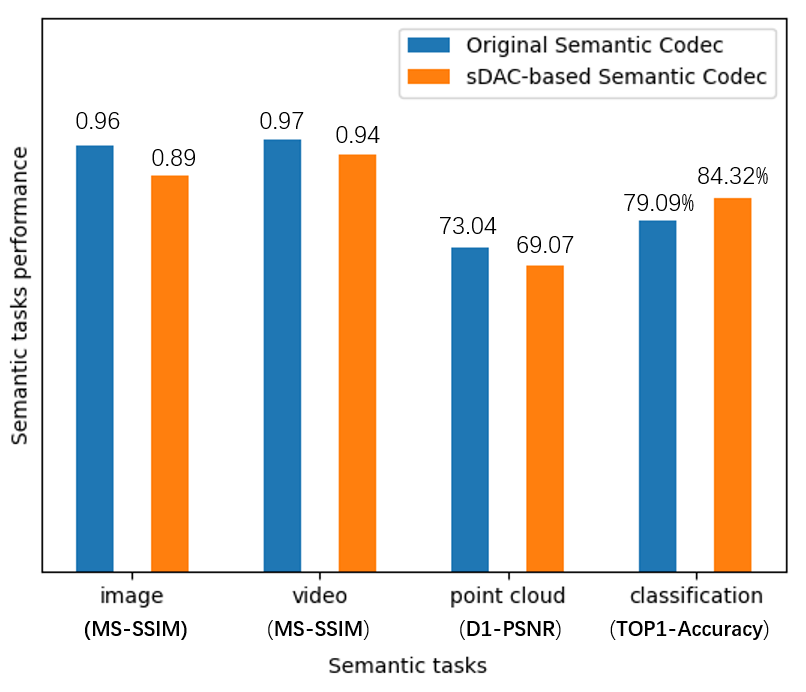}}
  \caption{Results of different sDAC-based semantic communication tasks performance. It shows the great generality of sDAC. After integrating with sDAC, the original semantic communication system experienced a slight performance drop in the reconstruction semantic task. However, task-oriented semantic communication systems, such as image classification, show a performance improvement, indicating the potential of sDAC in future semantic task applications.
  \label{task}}
\end{figure}
For the validation of sDAC's compatibility, sDAC is integrated into various semantic communication systems for various semantic tasks, and the results are shown in Fig.~\ref{task}. As different semantic tasks are measured by different metrics of different ranges, data in Fig.~\ref{task} is disproportional for clearance. The performance of the image and video is calculated using MS-SSIM, as mentioned above. Point cloud communication system is measured by D1-PSNR, and classification is measured by top-1 accuracy. Data left is the performance of the original semantic communication system, and data right is the performance of the original semantic communication system integrated with sDAC for compatibility with digital communication. It can be found that in traditional data transmission semantic tasks, sDAC will encumber the end-to-end performance a little bit because of the quantization error. However, in rising task-oriented semantic communication systems, such as classification semantic tasks, the integration of sDAC can quantize task-related data into discrete bits without an evident performance drop, indicating its potential in future task-oriented semantic communication systems.

\section{Conclusion}  
The goal of this paper is to propose a novel quantization and bits conversion algorithm for the compatibility of semantic communications and digital communications. Based on this, we propose the semantic digital analog converter (sDAC) to realize the bi-directional conversion of continuous values and binary discrete bits. It utilizes a pair of quantization adapters, a learnable codebook and an abstract BSC model to realize a generalized and noise-robust communication module. The quantization adapter makes it feasible to be compatible with lots of semantic communication systems in random quantization order. Learnable codebook studies the data distribution divergence between the semantic codec's data and quantized data to narrow down the kl divergence. The abstract BSC model efficiently simulates the effect caused by different digital communication schemes, enabling sDAC to be a great generality to various modulation methods. In the end, the performance of sDAC is validated across different semantic models, semantic tasks, modulation methods, channel conditions and quantization orders. In all experiments, sDAC shows great generative and channel robust properties compared with other communication schemes.

% or


% \appendices

% \section{}

% \section*{Acknowledgment}


\ifCLASSOPTIONcaptionsoff
  \newpage
\fi



\bibliographystyle{IEEEtran}


% \end{thebibliography}

% \small
\bibliography{IEEEabrv, reference}


% \end{IEEEbiography}


%\newpage



%\vfill


\end{document}